\documentclass[aps,pra,showpacs,twocolumn]{revtex4-1}
\usepackage{bm,graphicx}
\usepackage{bbm}
\usepackage{amssymb,amsmath,amsfonts,amsthm}
\usepackage[mathscr]{eucal}
\usepackage{color}

\usepackage[normalem]{ulem}

\usepackage[caption=false]{subfig}

\newcommand{\tr}{{\rm tr}\,}
\newcommand{\ket}[1]{\left|{#1}\right\rangle}

\newcommand{\braket}[2]{\langle{#1}|{#2}\rangle}
\newcommand{\ketbrad}[1]{\left|{#1}\rangle\!\langle{#1}\right|}

\begin{document}

\title
{Quantitative bound entanglement in two-qutrit states}
\author{Gael Sent\'{\i}s$^{1}$, Christopher Eltschka$^2$, Jens Siewert$^{3,4}$}
\affiliation{
$^{1}$Departamento de F\'{i}sica Te\'{o}rica e Historia de la Ciencia, Universidad del Pa\'{i}s Vasco UPV/EHU, E-48080 Bilbao, Spain\\
$^{2}$Institut f\"ur Theoretische Physik, Universit\"at Regensburg, D-93040 Regensburg, Germany\\
$^{3}$Departamento de Qu\'{i}mica F\'{i}sica, Universidad del Pa\'{i}s Vasco UPV/EHU, E-48080 Bilbao, Spain\\
$^{4}$IKERBASQUE Basque Foundation for Science, E-48013 Bilbao, Spain
}

\begin{abstract}

Among the many facets of quantum correlations,
bound entanglement has remained one the most enigmatic phenomena,
despite the fact that it was discovered in the early days of
quantum information. Even its detection has proven to be difficult,
let alone its precise quantitative characterization.
In this work, we present the exact quantification of entanglement
for a two-parameter family of highly symmetric two-qutrit mixed states,
which contains a sizable part of bound entangled states. We achieve
this by explicitly calculating the convex-roof extensions of the
linear entropy as well as the concurrence for every state within
the family. Our results provide a benchmark for future quantitative
studies of bipartite entanglement in higher-dimensional systems.
\end{abstract}


\maketitle

\emph{Introduction}.
Entanglement is considered one of the central resources
to perform tasks in quantum-information processing. 
In a realistic setting, the quantum 
states---as the carrier of this resource---are more or less mixed.
Therefore it is of high practical relevance that entanglement can
be  {\em distilled} to its pure form occurring in
singlet states~\cite{Bennett1996}. It came as a surprise when
Horodecki {\em et al.} discovered that there
exists so-called {\em bound entanglement} from which it is not possible
to distill even a tiny fraction of singlet entanglement
by means of local operations and classical communication~\cite{Horodecki1998}.

Technically, bound entanglement was found by investigating the
properties of the partial
transpose of $d_A\times d_B$-dimensional bipartite quantum states.
Whereas for dimensions $2\times 2$ and $2\times 3$ positivity of the
partial transpose (PPT) is sufficient for 
separability of the state~\cite{Horodecki1996a,Peres1996},
for higher local dimensions such as $3\times 3$ and $2\times 4$
(but also for three qubits $2\times 2\times 2$) there exist states
which are both entangled and PPT, and therefore
bound entangled. In continuous variables systems, the PPT criterion also suffices to reveal separability for bipartite mode Gaussian states with $1\times n$ modes~\cite{Werner2001a}. We mention that it is not known
whether there are bound entangled states with non-positive partial transpose (NPT).

While to date many aspects of entanglement are quite thoroughly 
understood both at a qualitative and a quantitative 
level~\cite{HorodeckiRMP2009,ESReview2014}, PPT entanglement 
continues to play an elusive role. 
Geometrically, PPT-entangled
states are located in close proximity to the separable states,
and hence were thought to be useless for quantum information processing. 
Remarkably, 
 it was established that they
may serve as a resource, e.g., for 
quantum-key distribution as well as 
in entanglement activation~\cite{Horodecki2005,Horodecki1999a,Masanes2006}. 
Further, it turned out only very recently 
that PPT-entangled states may allow for steering~\cite{Moroder2014a} and 
nonlocality~\cite{VertesiBrunner2014}. On the other hand,
there is no 
systematic theory on PPT entanglement, and
many interesting problems have remained open. Examples are 
the questions whether PPT-entangled states allow for entanglement
swapping~\cite{Bauml2010}, and what is their maximum possible
Schmidt number~\cite{Sanpera2001}.
 
One of the main reasons for the difficulty in systematically
studying the physics of PPT entanglement
is the lack of quantitative methods in the framework of
established entanglement theory. There exist numerical lower 
bounds for certain families of states, e.g., 
Refs.~\cite{Mintert2005,Otfried2012,Moroder2014},
and an analytical lower bound based on the computable cross
norm (or realignment) criterion~\cite{Chen2005}.
However, to our knowledge, there is no example of exact 
entanglement quantification for PPT-entangled states. Our work 
takes a step in this direction through the exact calculation
of the linear entropy and the concurrence~\cite{Albeverio2001,Rungta2001} 
for a subfamily of highly symmetric mixed two-qutrit states that 
display also PPT entanglement.
To this end, we first introduce a family of symmetric mixed $d\times d$
states by relaxing axisymmetry~\cite{Eltschka2013}. In particular we
study a subset of this family for two qutrits ($d=3$) that is 
parametrized by two real numbers. Subsequently, we obtain
the exact convex roofs for linear entropy and concurrence by combining 
established analytical and numerical methods~\cite{Osterloh2008,Moroder2014}.

\emph{The family}.
We characterize our family of interest as the set of bipartite states that remain invariant under some specific symmetry operations.
The family emerges naturally by relaxing the symmetries that define the two-parameter axisymmetric states introduced in  Ref. \cite{Eltschka2013}. Said symmetries are the ones to be found on the maximally entangled state \mbox{$\ket{\Phi}=(1/\sqrt{d})\sum_{j=0}^{d-1} \ket{j}\otimes\ket{j}$}, namely $(a)$ subsystem permutation, $(b)$ simultaneous permutation of the basis elements of both parties, and $(c)$ simultaneous (local) phase rotations of the form $V(\varphi_1,\varphi_2,\ldots,\varphi_{d-1})=e^{i\sum_j \varphi_j \mathfrak{g}_j} \otimes e^{-i\sum_j \varphi_j \mathfrak{g}_j}$,
where $\mathfrak{g}_j$ are the diagonal generators of ${\rm SU}(d)$. We remove symmetry $(a)$ and relax symmetry $(b)$ by considering only \emph{cyclic} permutations of basis elements, and we restrict ourselves to real (symmetric) density matrices.

All bipartite qutrit states that follow this weaker version of the axisymmetry are fully described by three real parameters---we denote these states by $\rho^\diamond$. To see this, we reason as follows. First, the local rotations eliminate all off-diagonal elements which are not of the form $\rho^\diamond_{jj,kk}$. Second, hermiticity together with the cyclic permutation symmetry force elements $\rho^\diamond_{jj,kk}$ to be all equal, and restrict diagonal elements to three different types. Finally, taking into account that $\tr\rho^\diamond=1$ and $\rho^\diamond={\rho^\diamond}^*$, we are left with three real parameters. 

\begin{figure}
	\centering
	\includegraphics[width=\linewidth]{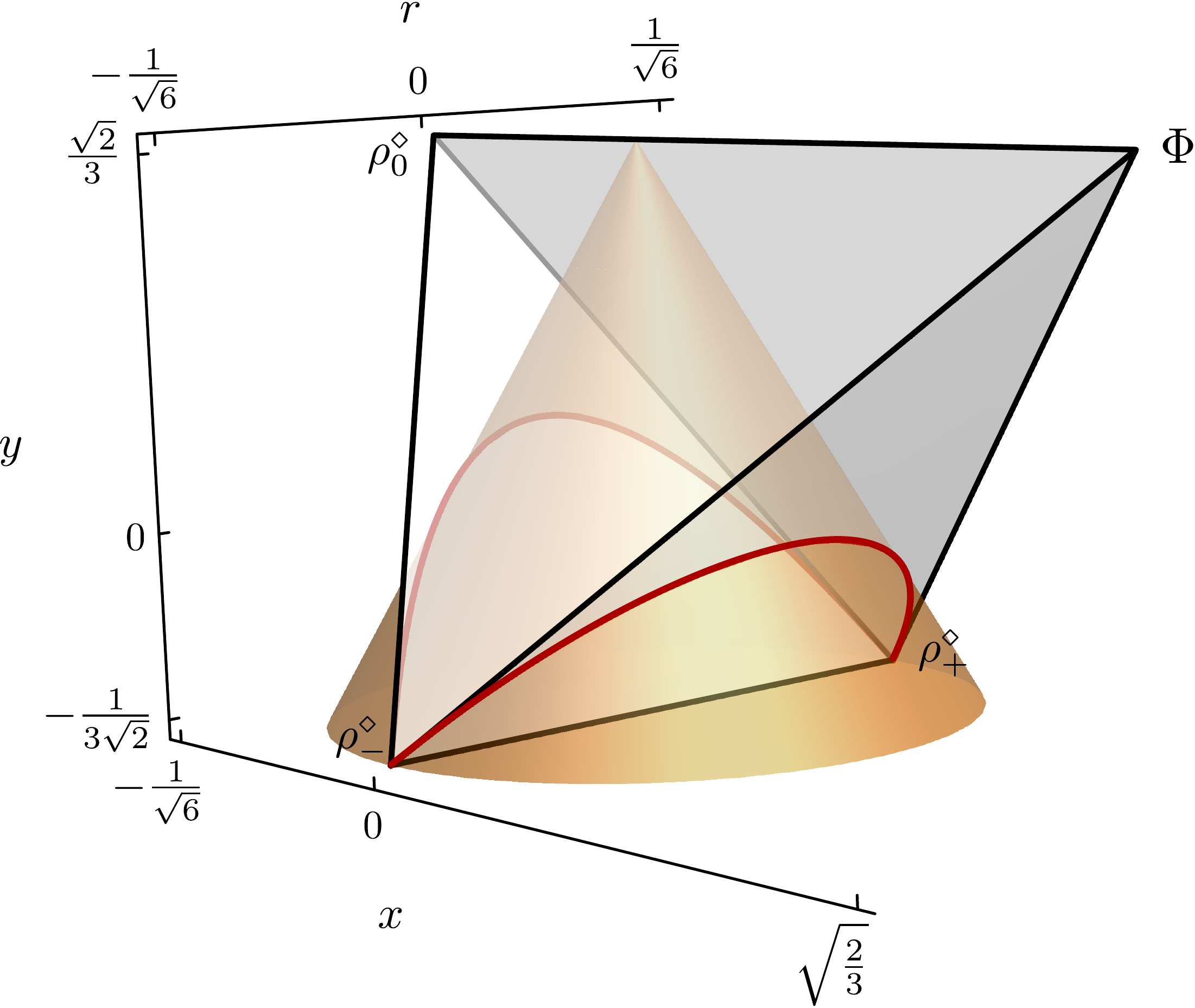}
	\caption{(Color online). The convex set of states $\rho^\diamond$, depicted as a tetrahedron in the space of parameters $x,y,r$. The four corners of the tetrahedron correspond to states $\Phi$, $\rho^\diamond_0$, $\rho^\diamond_{+}$ and $\rho^\diamond_{-}$. The boundary of the set of PPT states is a cone. The cone intersects two facets of the tetrahedron (red lines) and is tangent to the other two.
The exact calculation of entanglement measures is done for
the states on the lower-right facet of the tetrahedron.
}
	\label{fig:tetrahedron}
\end{figure}

We parametrize the nonzero elements of $\rho^\diamond$ as: 
%
%
$\rho^\diamond_{jj,jj} = \frac{1}{9}+\alpha$, $\rho^\diamond_{jj_\pm,jj_\pm} = \frac{1}{9}-\frac{\alpha}{2}\pm \gamma$, and $\rho^\diamond_{jj,kk} = \beta$, where $j,k=0,1,2$ and $j_\pm=(j\pm 1)\!\!\!\!\mod 3$. We choose the scaling of parameters $\alpha$, $\beta$, and $\gamma$ such that the distance between two points in the parameter space coincides with the Hilbert-Schmidt distance between the corresponding density matrices $D_{\rm HS}(X,Y)=\sqrt{\tr (X-Y) (X-Y)^\dagger}$. Thus, we switch to coordinates $(x,y,r)$ in a three-dimensional parameter space such that
$\alpha=y \sqrt{2}/3$, $\beta=x/\sqrt{6}$ and $\gamma=r/\sqrt{6}$. The positivity condition $\rho\geqslant0$ readily shows that the boundary of physical states is a tetrahedron, and limits the ranges of parameters as
$-1/\sqrt{6} \leqslant x \leqslant \sqrt{2/3}$, $-1/(3\sqrt{2}) \leqslant y \leqslant \sqrt{2}/3$, and $-1/\sqrt{6} \leqslant r \leqslant 1/\sqrt{6}$
(see Fig.~\ref{fig:tetrahedron}).

The intersection of the $r=0$ plane with the tetrahedron delimits the axisymmetric family, which contains the completely mixed state $\tfrac{1}{9}\openone_9$ at the origin and the Bell state $\Phi$ at the upper corner $(\sqrt{2/3},\sqrt{2}/3,0)$. 
The other upper corner $(-1/\sqrt{6},\sqrt{2}/3,0)$ corresponds to the mixture of Bell states 
$\rho^\diamond_0=(\Phi_++\Phi_-)/2$, where \mbox{$\ket{\Phi_\pm}=\tfrac{1}{\sqrt{3}}\sum_j e^{\pm i 2\pi(j-1)/3}\ket{jj}$} (note that, for brevity, 
we use the symbol
$\psi$ for the projector $\ketbrad{\psi}$).
The lower corner of the axisymmetric triangle is split in two, corresponding to the extremal states
$
\rho^\diamond_{\pm} = \frac{1}{3}\sum_{j} \ketbrad{jj_\pm}\,.
$
Note that all the relevant physical features of this family can be studied in the $r\geqslant 0$ region, since reversing the sign of $r$ is equivalent to 
permutation of the subsystems.

The PPT condition for states $\rho^\diamond$ simply reads \mbox{$\rho^\diamond_{jj,kk}\leqslant \sqrt{\rho^\diamond_{jj_+,jj_+}\rho^\diamond_{jj_-,jj_-}}$}. This is so because $(\rho^\diamond)^{T_A}$ only contains $2\times 2$ blocks, so that this upper bound on the off-diagonal elements is enough to ensure positivity of the eigenvalues.
In terms of $(x,y,r)$, the condition becomes $3\sqrt{6}\sqrt{x^2+r^2} \leqslant 2-3\sqrt{2}y$: it 
delimits a cone in the parameter space whose symmetry axis runs along the $y$ axis, and its base is tangent to states $\rho^\diamond_{\pm}$ (see Fig.~\ref{fig:tetrahedron}). The volume of the tetrahedron inside the PPT cone includes all the separable and the PPT-entangled states of the family.

Interestingly, the states $\rho^\diamond$ coincide with a family of $3\times 3$ states first introduced by Baumgartner \emph{et al.} \cite{Baumgartner2006} and further analyzed by Bertlmann 
and Krammer \cite{Bertlmann2007,*Bertlmann2008a,*Bertlmann2008,*Bertlmann2009}. 
In these works the family is constructed \emph{ad hoc} via the Weyl operator basis. However, the generating symmetries that we present here were not emphasized.


\emph{Linear entropy}.
The linear entropy of a pure state $\psi$ is given by $E_{\rm lin}(\psi)=2[(\tr\rho_A)^ 2-\tr(\rho_A^2)]$, where $\rho_A = \tr_A \ketbrad{\psi}$. As for any 
pure-state entanglement measure, this definition can be extended to mixed states via the convex roof~\cite{Uhlmann2010}
\begin{equation}\label{Elin_gen}
E_{\rm lin}(\rho) = \min_{\{p_k,\psi_k\}} \sum_k p_k E_{\rm lin}(\psi_k)\,,
\end{equation}
where the minimization is taken over all decompositions $\rho=\sum_k p_k \ketbrad{\psi_k}$ into pure states. While 
finding the optimal decomposition
for an arbitrary $\rho$ is generally a daunting challenge, the problem becomes tractable for families of states that exhibit certain symmetries~\cite{Vollbrecht2001}. This is the case of states $\rho^\diamond$. Albeit the techniques that we use are applicable to the whole family $\rho^\diamond$, in this paper we focus our attention on the subfamily of states 
\begin{equation}
   \sigma\ =\ z \ketbrad{\Phi}\ +\ 
              (1-z)\left[\frac{1+\bar{r}}{2} \rho^\diamond_+ + \frac{1-\bar{r}}{2}\rho^\diamond_-
                   \right] \,,
\label{eq:sigma}
\end{equation}
where $0\leqslant z\leqslant 1$ and $-1\leqslant \bar{r}\leqslant 1$. These are the states 
that lie on the lower facet of the tetrahedron, that is, the 
 triangle enclosed by states $\Phi$, $\rho^\diamond_{+}$, and 
$\rho^\diamond_{-}$
(cf.~Fig.~\ref{fig:tetrahedron}); we will extend the analysis to all states $\rho^\diamond$ elsewhere. 
Recall that, due to the $\bar{r}\leftrightarrow -\bar{r}$
symmetry, it suffices to study the parameter range $\bar{r}\in [0,1]$.
The boundary between PPT and NPT states in the $(\bar{r},z)$ parameter space is given by
\begin{equation}
         z_{\mathrm{boundary}}(\bar{r})\ =\ 
                                      \frac{-1+\bar{r}^2+2\sqrt{1-\bar{r}^2}}
                                      {3+\bar{r}^2}
\  \ .
\label{eq:sigma_PPTborder}
\end{equation}
%

We first calculate the linear entropy $E_{\rm lin}(\sigma)$ using the method of the convex characteristic curve \cite{Osterloh2008}, with $\sigma$ 
given in Eq.~\eqref{eq:sigma}.
The method is built upon the fact that the pure states in an optimal decomposition of $\rho$, i.e., one achieving the minimum in Eq.~\eqref{Elin_gen}, can be expressed as linear combinations of the elements of any other pure-state decomposition. Note that $\sigma$ is a rank-7 state, and that the (pure) elements $\psi_\sigma$ of its decompositions are restricted to live 
within its span. 
Thus an arbitrary state $\psi_\sigma$ can always be written as a linear combination of the vectors that span the facet, i.e.,
%
\begin{align}
\!\!\ket{\psi_\sigma} = \sqrt{z} \ket{\Phi} + \sqrt{1-z} 
\left[ \sqrt{(1+\bar{r})/2} \left(a\ket{01}+b\ket{20} \right.\right. \nonumber\\
 \left.+ c\ket{12}\right) + \left.\sqrt{(1-\bar{r})/2} \left(d\ket{10}+e\ket{02}+f\ket{21}\right) \right] \,, \label{psi_sigma}
\end{align}
where 
$\bar{r},z$ are the parameters that specify the corresponding symmetrized state $\sigma$ [cf. Eq.~\eqref{eq:sigma}], 
and $\{a,b,c,d,e,f\}\equiv\xi$ are 
complex free parameters subject to the normalization constraints $|a|^2+|b|^2+|c|^2=|d|^2+|e|^2+|f|^2=1$ [hence the number of real free parameters in Eq.~\eqref{psi_sigma} is 12, as it corresponds to an arbitrary state of dimension 7]. 

The first part of the method consists in computing the characteristic curve for the linear entropy $\tilde{E}_{\rm lin}(\bar{r},z)=\min_\xi E_{\rm lin}(\psi_\sigma)$ at each point in the facet.
We start with the expression of the linear entropy for an arbitrary bipartite pure state $\psi$ in terms of the state coefficients $E_{\rm lin}(\psi) =\sum_{jklm} \left|\psi_{jk}\psi_{lm}-\psi_{jm}\psi_{lk}\right|^2$, where $\psi_{jk}=\braket{jk}{\psi}$ for some orthonormal basis $\{\ket{jk}\}$. For the state $\psi_\sigma$ defined in Eq.~\eqref{psi_sigma}, 
the linear entropy is a function of the family coordinates $\bar{r}$ and $z$, and it depends on the free complex parameters $\xi$, i.e., $E_{\rm lin}(\psi_\sigma)=E_{\rm lin}(\bar{r},z,\xi)$.
The explicit form of this function immediately reveals that the complex phases of $\xi$ only play a role inside cosine multiplicative factors. Fixing all of them to the value $\pi$ minimizes each cosine factor independently, thus it suffices to consider real parameters. Taking this into account, the characteristic curve reads
%
\begin{align}\label{supp:cc}
\tilde{E}_{\rm lin}(\bar{r},z)=& \min_{\xi\in\mathbb{R}} \; \frac{4}{3} \Bigl\{  z -z(1-z)\sqrt{1-\bar{r}^2}(ad+be+cf) \nonumber\\
&-\sqrt{3z/2}(1-z)^{3/2}\left[ (1+\bar{r})\sqrt{1-\bar{r}}\right. \nonumber\\
&\times (bcd+ace+abf) \nonumber\\
&\left. +\sqrt{1+\bar{r}}(1-\bar{r})(cde+bdf+aef)\right]\nonumber\\
&+\frac{3}{4}(1-z)^2\left[(1+\bar{r})^2(a^2b^2+b^2c^2+a^2c^2)\right. \nonumber\\
&+(1-\bar{r}^2)(a^2d^2+b^2e^2+c^2f^2)\nonumber\\
&\left.+(1-\bar{r})^2(d^2e^2+e^2f^2+d^2f^2)\right]\Bigr\} \,.
%
%
\end{align}
%
Note that there are four free parameters left to minimize over (recall the normalization conditions $a^2+b^2+c^2=d^2+e^2+f^2=1$). Furthermore, a numerical evaluation of Eq.~\eqref{supp:cc} shows that, for all values of $\bar{r}$ and $z$, the minimum is still achieved if one sets $c=a$ and $f=d$, 
thus we can reduce the free parameters to only $\{b,e\}$. 
Although this is a great simplification of the original problem, the minimization turns out to be highly nontrivial, with at least three different regimes of solutions. An analytical formula for $\tilde{E}_{\rm lin}(\bar{r},z)$ hence becomes out of reach, so we resort to a numerical nonlinear multiparameter constrained minimization. As we will discuss, in principle this approach presents a problem of reliability that we nonetheless are able to circumvent.


The computed $\tilde{E}_{\rm lin}$ describes a nonconvex surface over the $(\bar{r},z)$ domain. The second part of the method is to compute its convexification $\tilde{E}^c_{\rm lin}$, which is, by construction, a lower bound of the true linear entropy of a generic (non-symmetric) state $\rho\in {\rm span}\{\ket{\psi_\sigma}\}\equiv\mathcal{S}$ \cite{Osterloh2008}. More precisely, at each point $(\bar{r},z)$ in the facet, the equation $\tilde{E}^c_{\rm lin}(\bar{r},z) \leqslant E_{\rm lin}(\rho)$ holds, where $\tr\rho\Phi = z$ and $\tr\rho\rho^\diamond_{+}=(1+\bar{r})(1-z)/6$. 
Here we go one step further and prove that, for a subset $\mathcal{K}\subset\mathcal{S}$ of states that are invariant under some entanglement-preserving symmetry, as the states $\sigma$ are, $\tilde{E}^c_{\rm lin}$ also represents an upper bound to their true linear entropy, hence we have the result
%
%
%
\begin{equation}\label{eq:ccc_is_exact}
\tilde{E}^c_{\rm lin}(\bar{r},z) = E_{\rm lin}(\sigma(\bar{r},z)) \, .
\end{equation}
That is, at this point, up to the reliability of our numerical calculation \footnote{Note that we have produced $\tilde{E}^c_{\rm lin}$ by means of a nonlinear constrained multiparameter minimization, for which no algorithm guarantees to find the global minimum. Were the minimization to fail at some of the explored coordinates $(\bar{r},z)$, the corresponding values $\tilde{E}^c_{\rm lin}(\bar{r},z)$ would upper bound the exact linear entropy.}, $\tilde{E}^c_{\rm lin}$ is the true linear entropy for the states on the facet. 

\begin{proof}[\indent Proof of Eq.~\eqref{eq:ccc_is_exact}]
We now proceed to prove that Eq.~\eqref{eq:ccc_is_exact} holds, independently of the chosen entanglement measure. Let $E$ be an entanglement measure and $\bar{r}, z, \ldots$ parametrize
the fidelities $\tr(\rho \rho_j)=f_j(\bar{r},z,\ldots )$ of some arbitrary state $\rho$ with some fixed given set of states
$\{\rho_j\}$.
While in general the value of the convex characteristic curve 
\begin{equation}
\tilde{E}^c(\bar{r},z,\ldots)\leqslant E(\rho)
\label{eq:lower}
\end{equation}
represents a lower bound to $E(\rho)$~\cite{Osterloh2008}, it
{\em coincides} with the true value $E(\sigma)$ 
\begin{equation}
   \tilde{E}^c(\bar{r},z,\ldots)\ =\ E(\sigma(\bar{r},z,\ldots))
\label{eq:equal}
\end{equation}
if the states $\sigma$ and $\{\rho_j\}$
are invariant under a group $G$ of entanglement-preserving symmetries, that
is, if $g \sigma g^{-1}=\sigma$,  $ g \rho_j g^{-1}=\rho_j$, 
and also $E(g \rho g^{-1})=E(\rho)$ for $g\in G$ and any $\rho$.

We first consider the case that $\tilde{E}^c(\bar{r},z,\ldots)=
E(\psi^c_{\sigma}(\bar{r},z,\ldots))$, i.e., the minimum value $\tilde{E}^c(\bar{r},z,\ldots)$ 
is realized by a single (not necessarily symmetric under $G$) pure state
$\psi^c_{\sigma}(\bar{r},z,\ldots)$ from the span of $\sigma$
with the same fidelity parameters. Now $\sigma(\bar{r},z,\ldots)$
can be obtained from
$\psi^c_{\sigma}(\bar{r},z,\ldots)$ simply by symmetrization with
respect to $G$
\begin{equation}\label{supp:symmetrized}
\sigma(\bar{r},z,\ldots) \ =\ \int dg\ g \psi^c_\sigma(\bar{r},z,\ldots) g^{-1}\ .
\end{equation}
Note that the fidelity parameters remain unchanged under symmetrization
if the states $\rho_j$ are $G$-invariant: 
$\tr\left[\left( g \psi_{\sigma}^c g^{-1}\right)\rho_j\right]
=\tr\left[\psi_{\sigma}^c \left(g^{-1}\rho_j g\right)\right]
=\tr\left[\psi_{\sigma}^c \rho_j \right]$. 
The symmetrization~\eqref{supp:symmetrized} corresponds
to an example pure-state decomposition of $\sigma(\bar{r},z,\ldots)$,
so that
\begin{align}
E(\sigma(\bar{r},z,\ldots)) & \leqslant  
             \int dg\  E\left(g \psi^c_\sigma(\bar{r},z,\ldots) g^{-1}\right)
\nonumber\\
            & =
             \int dg\  E\left(\psi^c_\sigma(\bar{r},z,\ldots) \right) 
\nonumber\\
            & =
             E\left(\psi^c_\sigma(\bar{r},z,\ldots) \right) 
\nonumber\\
            & =
             \tilde{E}^c(\bar{r},z,\ldots)\ \ .
\label{supp:Esymmetrized}
\end{align}
But since, according to our assumption,
$\tilde{E}^c(\bar{r},z,\ldots)$ {\em is} the minimum achievable
value for {\em any} decomposition of $\sigma(\bar{r},z,\ldots)$ [cf. Eq.~\eqref{eq:lower}], we must have
$E(\sigma(\bar{r},z,\ldots))=\tilde{E}^c(\bar{r},z,\ldots)$.

Alternatively, we may have the case that 
$\tilde{E}^c(\bar{r},z,\ldots)= \sum_k q_k E(\psi^c_{\sigma,k})$
with $\sum q_k=1$, i.e.,
the achievable minimum of $E$ at $\bar{r},z,\ldots$ 
results from a convex combination of pure states
$\psi^c_{\sigma,k}(\bar{r}_k,z_k,\ldots)$ within the span of $\sigma$.
While the
fidelity parameters $\bar{r}_k,z_k,\ldots$ are different from $\bar{r},z,
\ldots$, those of their convex combination are the same~\cite{Osterloh2008}:
$\tr\left[\left(\sum_k q_k\psi^c_{\sigma,k}\right)\rho_j\right]
=\tr\left[\sigma\rho_j\right]$.
Now, in analogy with the approach above, we may construct an 
example decomposition of $\sigma(\bar{r},z,\ldots)$ by symmetrization,
because $\sigma$ is uniquely defined by the fidelity parameters $\bar{r},z,
\ldots$:
\begin{equation}\label{supp:Ksymmetrized}
\sigma(\bar{r},z,\ldots) \ =\ \sum_k q_k \int dg\ g \psi^c_{\sigma,k}(\bar{r},z,\ldots) g^{-1}\ .
\end{equation}
Correspondingly, for the entanglement we have, by virtue of
the convexity of $E$,
\begin{align}
\label{supp:KEsymmetrized}
E(\sigma(\bar{r},z,\ldots)) & \leqslant  
             \sum_k q_k \int dg\  E\left(g \psi^c_{\sigma,k}(\bar{r},z,\ldots) g^{-1}\right)
\nonumber\\
            & =
             \sum_k q_k \int dg\  E\left(\psi^c_{\sigma,k}(\bar{r},z,\ldots) \right) 
\nonumber\\
            & =
             \sum_k q_k E\left(\psi^c_{\sigma,k}(\bar{r},z,\ldots) \right) 
\nonumber\\
            & =
             \tilde{E}^c(\bar{r},z,\ldots)\ \ .
\end{align}
Again, since $\tilde{E}^c(\bar{r},z,\ldots)$ already represents the
smallest possible value for those fidelity parameters the
equal sign must hold. Thus, we have shown that, for states that are invariant under the action of an entanglement-preserving group of symmetry operations
$G$, the convex characteristic curve not only gives a lower
bound to an entanglement measure, but there is always 
a decomposition that realizes this lower bound, confirming
that the convex characteristic curve coincides with 
the exact entanglement measure. We remark that, in case such invariance is lacking, the convex characteristic curve holds only as a lower bound to the entanglement measure.
\end{proof}


%
\begin{figure}
	\centering
	\includegraphics[width=\linewidth]{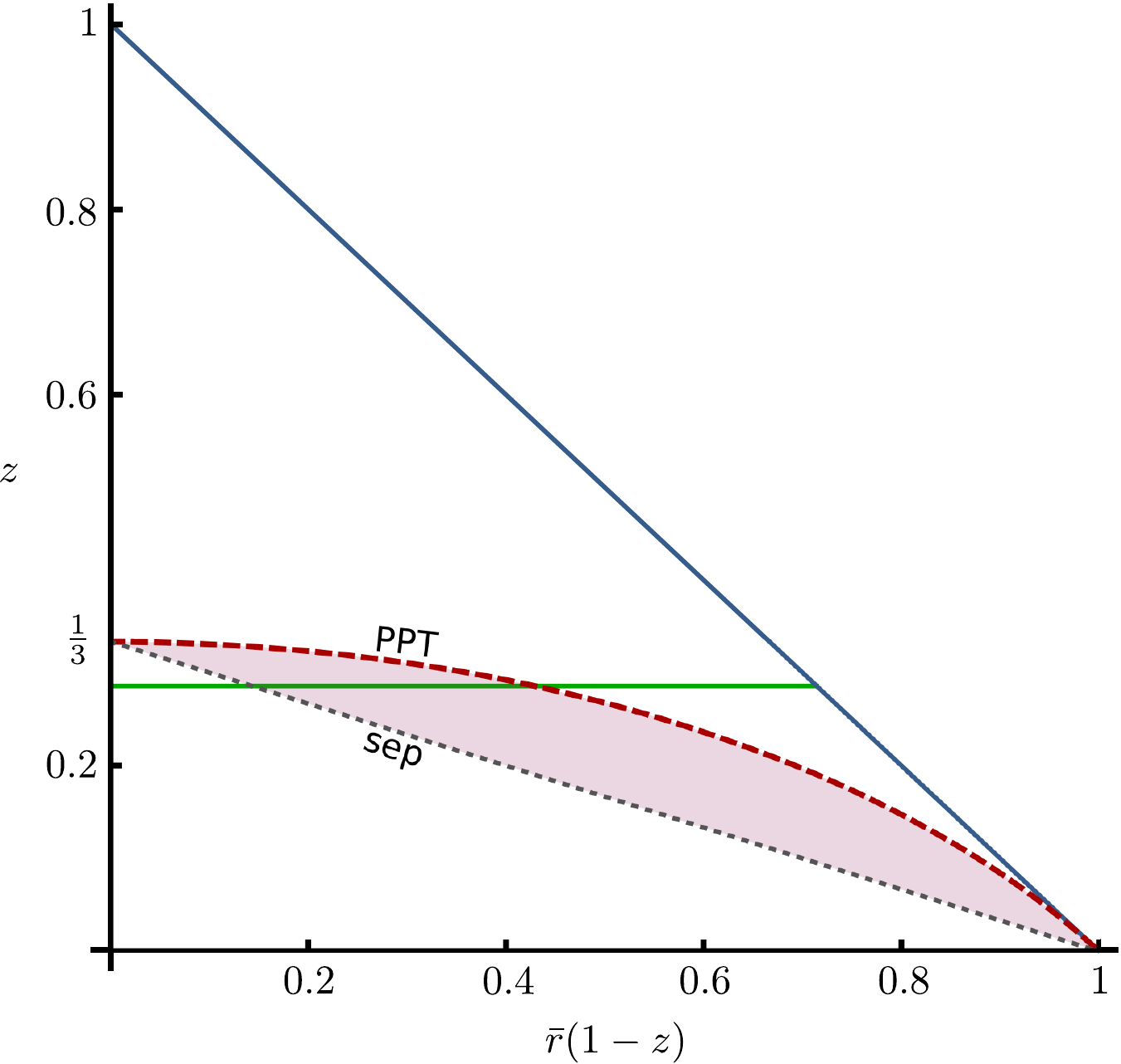}
	\caption{(Color online). The convex set of states $\sigma$, 
i.e., those that belong to the lower facet of the tetrahedron (cf.
Fig.~\ref{fig:tetrahedron}), for $\bar{r}\geqslant 0$. 
The border of the set of separable states is given 
by the straight line $z=1/3[1-\bar{r}(1-z)]$ (dotted gray line), 
while the boundary between PPT and 
NPT states, Eq.~\eqref{eq:sigma_PPTborder},
 is represented by the dashed red line. 
The filled region corresponds to all states $\sigma$ that are PPT entangled. 
The border of the triangle (solid blue line) delimits the set of physical states.  
The solid horizontal line (green) 
at $z=2/7$ represents the one-parameter
family of $3\times 3$ bound entangled states of Horodecki 
{\em et al.}~\cite{Horodecki1999a}.  }
	\label{fig:triangle}
\end{figure}

In addition to the numerical treatment of $\tilde{E}^c_{\rm lin}$ described above, we note that it is actually possible to characterize analytically the set of separable states.  For $\tilde{E}_{\rm lin}(\bar{r},z)=0$ one can find 
an analytical condition in terms of a higher-order polynomial.
The resulting zero line contains the separable points
at $(\bar{r},z)=(0,1/3)$ and $(1,0)$ and 
lies below the straight line $z=1/3[1-\bar{r}(1-z)]$. Since 
the set of separable states has to be convex, its border must 
be given by that straight line
(cf.~Fig.~\ref{fig:triangle}). Alternatively, one could use the 
computable cross norm to obtain the same result~\cite{Bertlmann2008,Bertlmann2009}.

\begin{figure}[h]
	\centering
	\includegraphics[width=.49\textwidth]{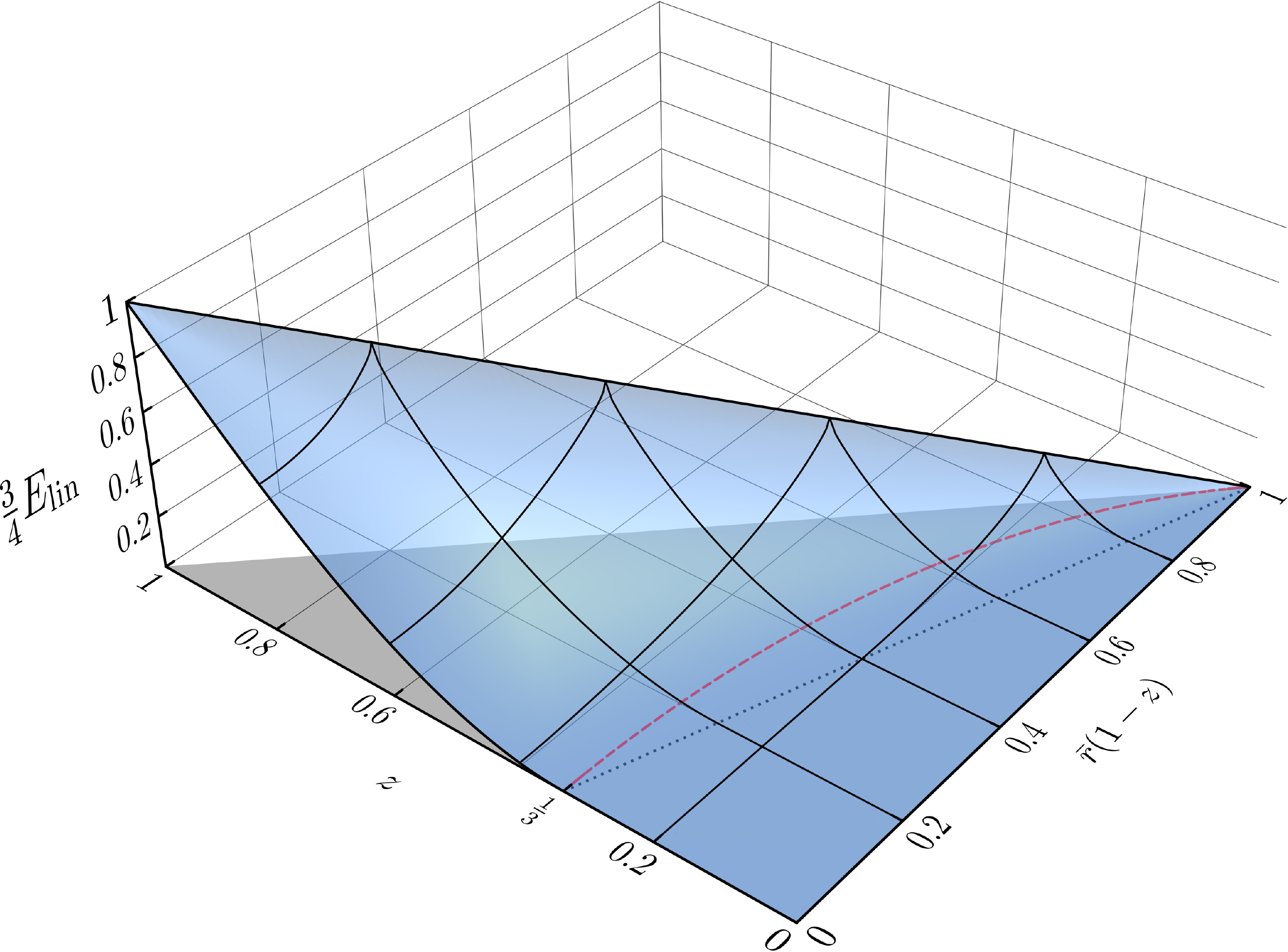}	
\caption{(Color online). Exact convex roof for the 
          linear entropy $\tilde{E}^c_{\rm lin}$ of the family 
          $\sigma$, Eq.~\eqref{eq:sigma}. Notably, the boundary between
        PPT and NPT states [red solid line in the $(\bar{r},z)$ plane]
        is not at all special for the dependence of the linear 
        entropy on the parameters.}\label{fig:Elin}
\end{figure}

To finish this section, we now compute a lower bound to the linear entropy using a recent method proposed in~\cite{Moroder2014} and show that it coincides with $\tilde{E}^c_{\rm lin}$, thereby certifying our numerical minimization. The method is based on semidefinite programming (SDP)~\cite{Vandenberghe1996}, and is able to produce a lower bound for the convex roof extension of any measure that, for pure states, can be written as a low-order polynomial of operator expectation values. Given a test state $\rho\in\mathcal{H}$, the method requires that the measure is cast as a suitable operator acting on a symmetric extension $\omega_{12\ldots N}\in\mathcal{H}^ {\otimes N}$, with $N$ large enough for the output to be polynomial. Then, an SDP minimization of the measure is carried out over the variable $\omega_{12\ldots N}$ subject to a (necessary) separability constraint, i.e., the PPT criterion \cite{Peres1996,Horodecki1996a}. This is in the same spirit as the PPT-symmetric extensions hierarchy introduced 
by Doherty {\em et al.}~\cite{Doherty2004,*Doherty2005}. 

Fortunately, the linear entropy becomes a polynomial with only a two-copy extension \cite{Horodecki2003}, hence  the dimension of $\omega_{12}$ is limited to 17 (that is the dimension of the symmetric subspace of two copies of $3\times 3$ states) and the SDPs are solved efficiently. 
The result is shown in Fig.~\ref{fig:Elin}. 
At this point it is worth stressing that the surface $\tilde{E}^c_{\rm lin}$ is obtained numerically from the convex hull of a finite set of three dimensional points, i.e., those where we have evaluated $\tilde{E}_{\rm lin}(\bar{r},z)$. Thus $\tilde{E}^c_{\rm lin}$ is not a smooth surface, for it depends on the chosen resolution of the grid of points. This is the reason behind the observed discrepancy of order $10^{-5}$ at worst between the SDP-computed lower bound and $\tilde{E}^c_{\rm lin}$, for a grid of $\sim10^5$ points. Remarkably, the discrepancy for $\sim 90\%$ of the grid is below $10^{-9}$.

\emph{Concurrence}.
Next, we show how the calculations we have done so far provide us also with the exact concurrence of states $\sigma$.
The concurrence of a pure state $\psi$ is directly related to its linear entropy by $C(\psi)=\sqrt{E_{\rm lin}(\psi)}$. Hence the characteristic curve of the concurrence for the facet is simply $\tilde{C}(\bar{r},z)=\min_\xi \sqrt{E_{\rm lin}(\psi_\sigma)} = \sqrt{\tilde{E}_{\rm lin}(\bar{r},z)}$. Since the SDP bound endorses $\tilde{E}^c_{\rm lin}$ as the exact linear entropy, in turn derived from $\tilde{E}_{\rm lin}$, we  conclude 
that $\tilde{C}$ is as good. In the same vein as before, its convexified version $\tilde{C}^c$ coincides with the exact concurrence \cite{Osterloh2008}. 
It turns out that $\tilde{C}^ c$ is much simpler than $\tilde{E}^c_{\rm lin}$. 
The concurrence of the states at points (0,1/3) and (1,0) (diagonal, thus separable) as well of the state at (0,1) (the maximally entangled state) are known, and define a plane that gives an upper bound to $\tilde{C}^c$. As it turns out, $\tilde{C}$ nowhere is below that plane, therefore said plane is indeed the convexified function in that region. This is summarized by the simple formula
$
\tilde{C}^c(\bar{r},z) = [(3-\bar{r})z-(1-\bar{r})]\sqrt{3}
$ for $z \geqslant (1-\bar{r})/(3-\bar{r})$, and $\tilde{C}^c(\bar{r},z) = 0$ otherwise.
%



{\em Discussion and outlook}.
We have presented a family of highly symmetric $d \times d$
bipartite mixed states for which we have specifically 
studied the case of two qutrits ($d=3$). We have exploited the symmetry of the
states to calculate the exact linear entropy for the states $\sigma$
in a two-parameter
subfamily, Eq.~\eqref{eq:sigma}, which includes also PPT-entangled
states. To this end, we have employed a combination of the 
convex characteristic curve~\cite{Osterloh2008} and an SDP
method~\cite{Moroder2014}. Moreover, based on the {\em a posteriori} 
certification of the straightforward numerical minimization (yet without 
analytical proof) we obtain also
the concurrence for the same subset of states as a simple affine function.
These results display several remarkable features. First, we note that
neither the linear entropy nor the concurrence exhibit any peculiarity
in their behavior along the boundary, Eq.~\eqref{eq:sigma_PPTborder}, 
between PPT and NPT-entangled states.
This may be viewed as an indication that the resources quantified
by those entanglement measures do not bear a direct relation to the
concept of distillability. Apart from that, the functional dependence of
the concurrence---as a polynomial entanglement measure of homogeneous
degree 1 in the density matrix---is the simplest one possible: an affine
function. An interesting technical point is that the SDP method for the 
linear entropy, in principle, uses a hierarchy in the number of symmetric
extensions~\cite{Moroder2014}. Notably, our work shows that for the states 
$\sigma$ in Eq.~\eqref{eq:sigma}
the exact solution is obtained already at the first level of the hierarchy.  
Last but not least it is worth mentioning that for the 
one-parameter family of $3\times 3$
bound-entangled states described by Horodecki {\em et al.}~\cite{Horodecki1999a}
there are numerous studies in the literature.
Our exact result now provides a firm
ground from quantitative entanglement theory for those investigations.

Apart from the obvious extensions of this work, 
such as an investigation of the interior of the $3\times 3$ tetrahedron,
the most interesting aspect for PPT entanglement, and bipartite entanglement
in general, results from the following observation. The first exact 
characteristics of $d\times d$ bipartite entanglement were obtained for
isotropic states~\cite{Horodecki1999}. 
Relaxing that symmetry leads to the axisymmetric states
which are still quite similar to the isotropic ones, and allow for 
exact treatment~\cite{Eltschka2015a}. Relaxing, in turn, the axisymmetry
as in the present study includes also PPT entanglement in the properties
amenable to an analysis without approximations. 
That is, the relaxed axisymmetry provides
the grounds for a systematic investigation of PPT entanglement
and thus for a deeper insight into the structure of the state space
of bipartite systems.

\emph{Acknowledgments}.
This work was funded by the German Research Foundation within 
SPP 1386 (C.E.), by Basque Government Grant No. IT-472-10,
MINECO Grants No. FIS2012-36673-C03-01  and No. FIS2015-67161-P, by EU ERC Starting Grant No. 258647/GEDENTQOPT,
and UPV/EHU program UFI 11/55 (G.S.\ and J.S.). 
The authors would like to thank O.\ G\"uhne, M.\ Huber, R.\ Mu\~noz Tapia, 
A.\ Monr\`as, G.\ T\'oth, and T.\ V\'ertesi for stimulating discussions
and comments.
%

\bibliography{Jens_Gael_QBasq_noarxiv}

\end{document}